\documentclass[pra,twocolumn,preprintnumbers,amsmath,amssymb]{revtex4}
\usepackage{graphicx}
\usepackage{dcolumn}
\usepackage{bm}
\usepackage{amsmath}

\DeclareGraphicsExtensions{.pdf,.eps,.png,.jpg,.mps}

\usepackage[utf8]{inputenc} 

\begin{document}

\title{Stacking change in MoS$_{2}$ bilayers induced by interstitial Mo impurities}

\author{Natalia Cort\'es}
\email{natalia.cortesm@usm.cl}
\affiliation{Departamento de F\'{i}sica, Universidad T\'{e}cnica Federico Santa Mar\'{i}a,
Casilla 110V, Valpara\'{i}so, Chile}

\author{L. Rosales}
\affiliation{Departamento de F\'{i}sica, Universidad T\'{e}cnica Federico Santa Mar\'{i}a,
Casilla 110V, Valpara\'{i}so, Chile}

\author{P. A. Orellana}
\affiliation{Departamento de F\'{i}sica, Universidad T\'{e}cnica Federico Santa Mar\'{i}a,
Casilla 110V, Valpara\'{i}so, Chile}

\author{A. Ayuela}
\affiliation{Centro de F\'{i}sica de Materiales (CSIC-UPV/EHU)-Material Physics Center (MPC), Donostia International Physics Center (DIPC), Departamento de F\'{i}sica de Materiales, Fac. Qu\'{i}micas UPV/EHU. Paseo Manuel de Lardizabal 5, 20018, San Sebasti\'an-Spain.}

\author{J. W. Gonz\'alez}
\affiliation{Centro de F\'{i}sica de Materiales (CSIC-UPV/EHU)-Material Physics Center (MPC), Donostia International Physics Center (DIPC), Departamento de F\'{i}sica de Materiales, Fac. Qu\'{i}micas UPV/EHU. Paseo Manuel de Lardizabal 5, 20018, San Sebasti\'an-Spain.}

\date{\today}

\begin{abstract}
We use a theoretical approach to reveal the electronic and structural properties of molybdenum impurities between MoS$_{2}$ bilayers.
We find that interstitial Mo impurities are able to reverse the well-known stability order of the pristine bilayer, because the most stable form of stacking changes from AA'(undoped) into AB (doped).
The occurrence of Mo impurities in different positions shows their split electronic levels in the energy gap, following octahedral and tetrahedral crystal fields.
The energy stability is related to the accommodation of Mo impurities compacted in hollow sites between layers.
Other less stable configurations for Mo dopants have larger interlayer distances and band gaps than those for the most stable stacking.
Our findings suggest possible applications such as exciton trapping in layers around impurities, and the control of bilayer stacking by Mo impurities in the growth process.

\end{abstract}


\maketitle
\date{Today}
\section{Introduction} \label{Intro}

The recent isolation of new 2D materials, such as hexagonal boron nitride (h-BN) \cite{novoselov2005two}, black phosphorus \cite{castellanos2014isolation}, and particularly transition metal dichalcogenides (TMDCs) \cite{yang1991structure} have attracted considerable attention thanks to their interesting physical, chemical, electronic, optical, and mechanical properties \cite{splendiani2010emerging,chhowalla2013chemistry,wang2012electronics,PhysRevB.88.245436,radisavljevic2011single,oriol2013,Hualing2012,castellanos2012}.
Among TMDCs,  MoS$_{2}$ is being used as prototype in several different applications, such as photovoltaic cells, photocatalysts, electronic nanodevices, and energy storage and conversion materials \cite{Book1}.
In structural terms, a layer of MoS$_{2}$ has Mo centers coordinated with six sulfur ligands in a trigonal prismatic arrangement \cite{liu2015electronic,roldan2014electronic}, following a hexagonal lattice of alternating Mo and S atoms, as seen from above.
TMDCs nanostructures can then be produced by stacking several MoS$_2$ layers through weak van der Waals interactions. It is noteworthy that the number of layers and their stacking arrangement largely modify the electronic properties of the MoS$_2$ semiconductor \cite{ataca2011comparative,lopez2016band,yuan2016evolution,he2014stacking,yan2016identifying}.
For instance, the MoS$_{2}$ monolayer shows a direct band gap, compared with the indirect band gap of MoS$_{2}$ bulk  \cite{mahatha2012electronic,zhao2013origin,kosmider2013electronic,padilha2014nature}.
The stacking in the MoS$_{2}$ bilayer alters the band gap, which can then be engineered not only by strain \cite{dong2014theoretical,sharma2014strain}, but  also by sliding \cite{levita2014sliding} and twisting the MoS$_{2}$ layers \cite{liu2014evolution}.
Thus, the control of the MoS$_{2}$ bilayer stacking is relevant when considering different applications in future devices.

\begin{figure}[t]
\centering
\includegraphics[clip,width=\columnwidth]{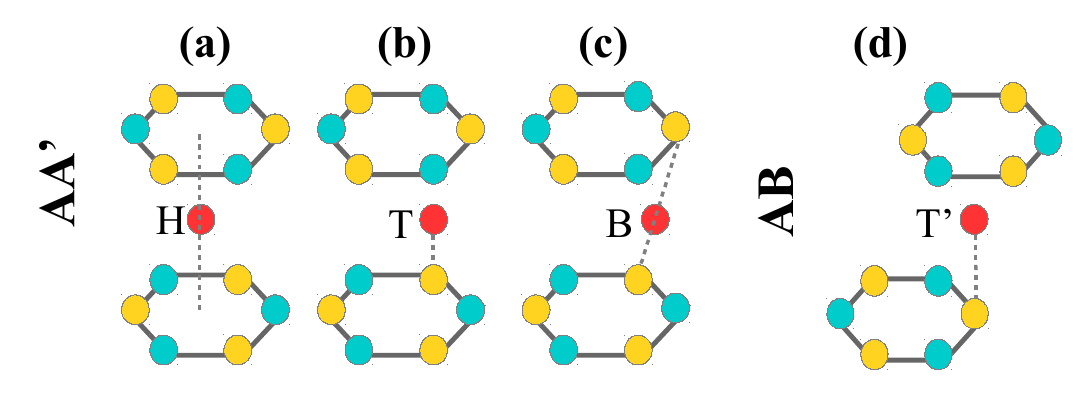}
\caption{Representation of a MoS$_{2}$ bilayer with two types of stackings, AA'and AB, with the Mo impurity atom in different positions. Mo atoms are shown as cyan spheres, sulfur atoms are shown as yellow spheres, and the Mo impurity as red spheres. (a-c)  AA' stacking with three initial positions for the Mo impurity atom: (a) hollow (H), (b) top (T), and (c) bridge (B). (d) AB stacking with the Mo impurity atom in the top position (T').}\label{fig1}
\end{figure}

The electronic and magnetic properties of MoS$_{2}$ monolayers and bilayers are also tuned by defects, such as vacancies and adatoms \cite{chang2014atomistic,huang2013density,wang2015electronic}. Most of the adatom impurities considered within MoS$_{2}$ layers are in groups  IA and VIIA, or transition metal atoms. The adatoms in the interlayer space have interesting effects, such as \textit{n}- or \textit{p}-type doping, induced magnetic moments \cite{wang2014first,he2010magnetic,lu2014electronic}, and structural phase transitions \cite{pandey2016phase}.
Although some early photoelectron spectroscopy experiments appeared to show Mo atoms embedded between MoS$_{2}$ layers \cite{fives1992photoelectron},
current experiments using low-temperature STM show Mo impurities between bilayers \cite{wan2013incorporating}.
However, the effects of intrinsic Mo impurity atoms on the electronic and structural properties of MoS$_{2}$ bilayers with different stackings are still largely unexplored.

We present a theoretical study of the structural and electronic properties of the MoS$_{2}$ bilayer considering Mo atoms as intrinsic impurities, placed at different positions within the interlayer region in a diluted regime. Using density functional with van der Waals  calculations from first principles, we relax the structures and study their electronic properties. We find that the intrinsic Mo impurities in the interlayer region produce some interesting behaviors, namely (i) a change in the stability order with respect to the pristine bilayer, energetically favoring the AB stacking over the AA', (ii) impurity states in the band gap region, and (iii) an increase in the distance between layers.
The structural and electronic modifications induced by the impurities could be employed as electron and exciton-like traps \cite{he2013experimental,castellanos2013local}, and the change in stacking produced by the Mo impurities could also be useful to fine-tune the stacking during the growing process \cite{liu2014evolution,PhysRevB.93.041420}.

\section{Model and Computational Details} \label{compdetail}
A Mo doped MoS$_2$ bilayer is described using density functional of van der Waals (vdW-DF) calculations with the SIESTA method \cite{soler2002siesta}. To describe the core electrons, we consider norm-conserving  relativistic ab-initio pseudopotentials in the Troullier Martins form \cite{PhysRevB.43.1993},  including nonlinear core corrections for inner d-electrons \cite{PhysRevB.26.1738}.
The exchange and correlation energy are calculated by the non-local vdW-DF, using the parametrization proposed by Dion \textit{et al} \cite{PhysRevLett.92.246401}, taking into account the exchange energy modification included by Cooper (C09) \cite{PhysRevB.81.161104}.
To include impurities, the structures of the MoS$_{2}$ bilayers are extended to a $3\times3$ MoS$_{2}$ supercell, using periodic boundary conditions. A \textit{k}-grid of $10\times10\times1$ Monkhorst-Pack is used to sample the Brillouin zone. A vacuum region in the \textit{z}-direction, of at least 20 \AA{} avoids interactions with periodic images.
The structures were relaxed until the force in each atom was less than $10^{-2}$ eV/\AA.

We first consider the most stable configurations of the pristine MoS$_{2}$ bilayer, namely the AA' and the AB con\-fi\-gu\-ra\-tions.
In AA' stacking, the hexagons in each layer are superposed in such a way that the molybdenum atoms of the bottom layer are located just below the sulfur atoms in the top layer, and vice-versa. For AB stacking the hexagons in each layer are shifted with the sulfur atoms of the bottom layer beneath the hollow sites of top layer, and the molybdenum atoms in the top layer over the molybdenum atoms in the bottom layer \cite{he2014stacking}, as shown schematically in Fig. \ref{fig1}.
In our calculations we found an energy difference between these stackings of E$_\mathrm{AA'}$ - E$_\mathrm{AB}$ =  $2.6$ meV per atom, in good agreement with previous DFT calculations \cite{he2014stacking,yan2016identifying,yang2014vapor}.

We then include Mo atoms as intrinsic impurities within the interlayer region for both bilayer stackings at different inequivalent positions, as shown in Fig. \ref{fig1}. The Mo impurity is labeled as Mo$_\mathrm{{imp}}$. The initial absorption sites for Mo within the MoS$_{2}$ bilayer are assumed to follow Mo sites in MoS$_{2}$ monolayers \cite{komsa2015native,jinhua2017}.
For AA' stacking, we have (i) a hollow site (H) in the center of the two hexagonal rings formed by the Mo and sulfur atoms belonging to the MoS$_{2}$ bilayer, (ii) a top site (T) just above a sulfur atom of the bottom layer, and (iii) a bridge site (B) between two sulfur atoms of the MoS$_{2}$ bilayer. In the AB stacking we also included (iv) the top site (T') on the S atom.

\begin{figure*}[ht!]
\centering
\includegraphics[clip,width=\textwidth,angle=0,clip]{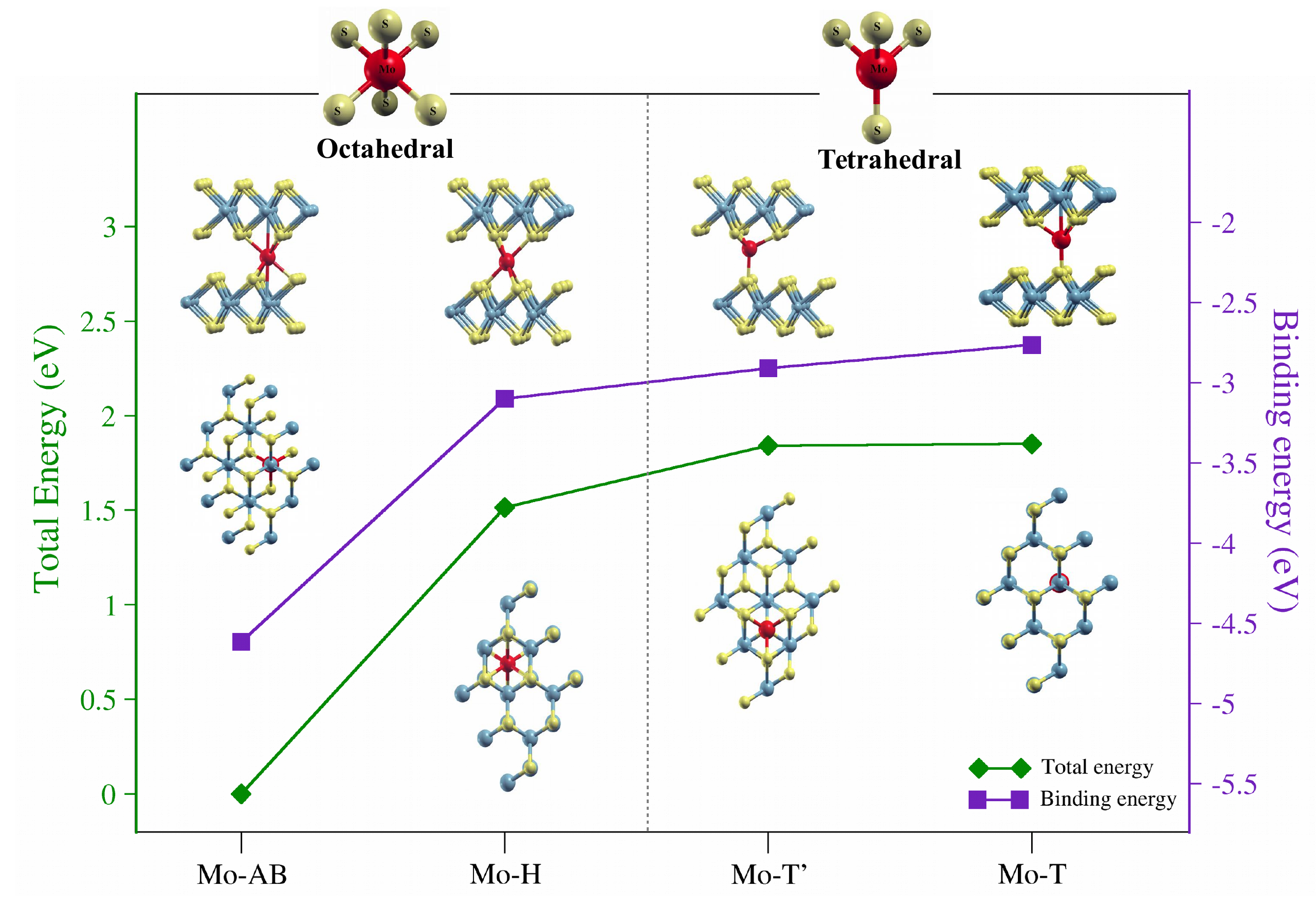}
\caption{Total and binding energy as functions of structural configurations of Mo impurities within the MoS$_2$ bilayer. The red spheres indicate the Mo impurity in each configuration. The relaxed structures are included, grouped in octahedral (Mo-AB and Mo-H) and tetrahedral (Mo-T' and Mo-T) structures of sulfur atoms around Mo impurities. The zero energy point is set for the most energetically favorable structure, namely the Mo-AB configuration. Figure prepared using XCrySDen \cite{xcrysden}.} \label{fig2}
\end{figure*}

\begin{figure*}[ht!]
\centering
\includegraphics[clip,width=\textwidth,angle=0,clip]{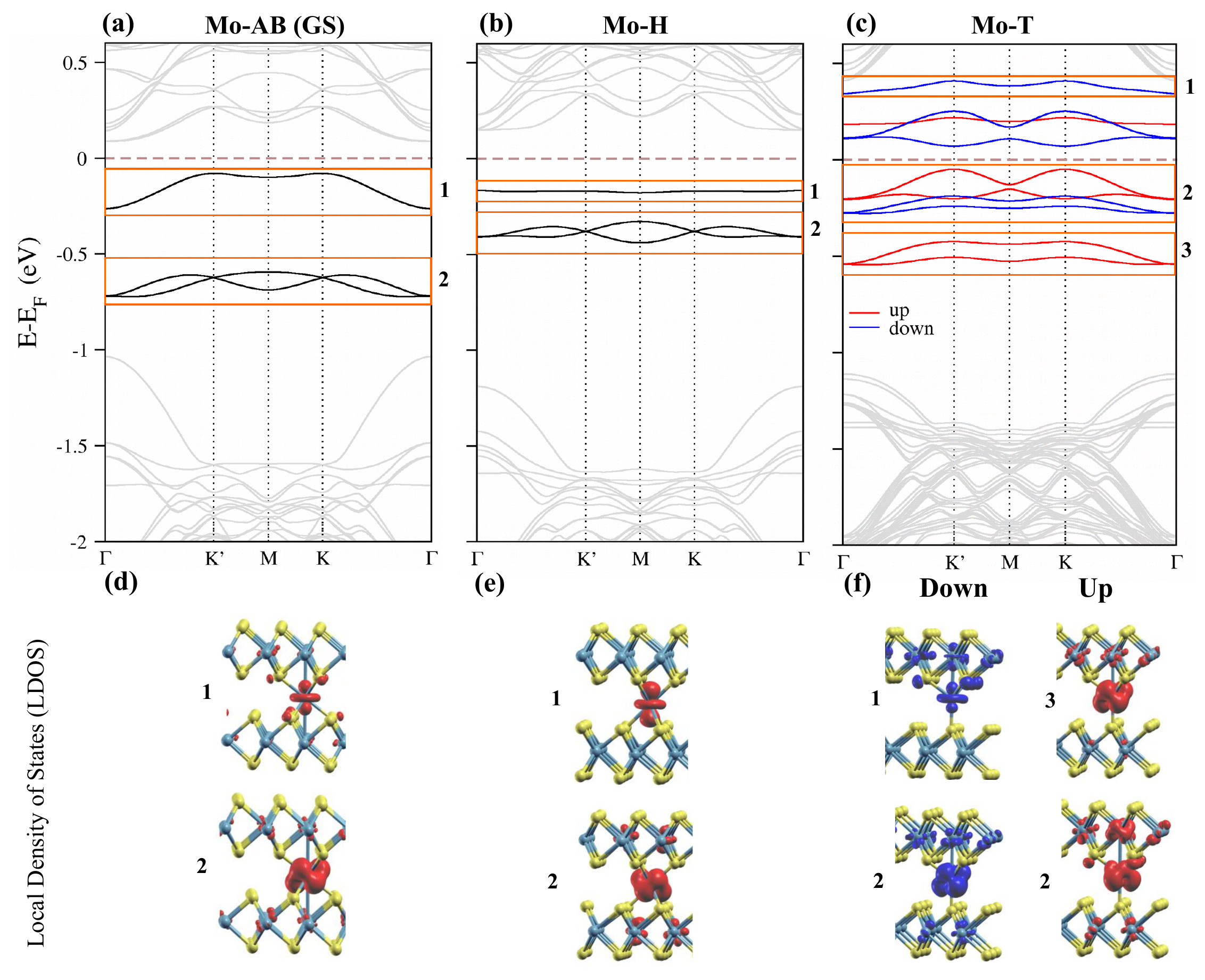}
\caption{(a-c) Band structures for the given configurations. The Fermi energy is set to 0 eV. Orange rectangles enclose the Mo$_\mathrm{imp}$ bands separated into several different energy regions, labeled as \textbf{1}, \textbf{2} and \textbf{3}. (d-f) Local density of states (LDOS) projected in space for the Mo$_\mathrm{imp}$ bands in the band gap region of the MoS$_{2}$ bilayer. Figure prepared using XCrySDen \cite{xcrysden}.} \label{fig3}
\end{figure*}

The MoS$_{2}$ bilayer structures with the impurity in the interlayer region are fully relaxed, to allow the optimized lattice parameters and atomic coordinates to be obtained. The binding energy can then be calculated using:
\begin{equation}
E_\mathrm{binding}=E_{\mathrm{Total}}-E_{\mathrm{bilayer}}-E_\mathrm{imp}\nonumber,
\end{equation}
where $E_\mathrm{{Total}}$ is the total energy of the MoS$_{2}$ bilayer with the impurity, $E_\mathrm{bilayer}$ is the energy of the corresponding final pristine MoS$_{2}$ bilayer (either AA' or AB), and $E_\mathrm{imp}$ is the energy for the isolated Mo$_\mathrm{imp}$ atom.

\section{Results and Discussion} \label{Resultados}

\subsection{Energy and Geometry}
Figure \ref{fig2} shows the total and binding energies for the different stackings and impurity positions.  The binding energies are negative, which indicates that the Mo impurity atoms are indeed adsorbed in the interlayer region of the MoS$_{2}$ bilayer.  The binding and total energies exhibit the same trend in terms of stability. The results in increasing order of stability shows that in the presence of the interlayer Mo$_\mathrm{imp}$, the Mo-AB bilayer configuration is the most energetically favorable. This configuration has AB stacking with the Mo$_\mathrm{imp}$ superposed with two Mo atoms as seen from above. Note that the Mo-AB configuration is reached from the input that has the Mo$_\mathrm{imp}$ placed at the bridge (B) position in the AA' stacking.

The next most favorable configuration is Mo-H, with the Mo$_\mathrm{imp}$ in the hollow position, in which the bilayer structure maintains the AA' stacking. The Mo-H case is less stable than Mo-AB by about $1.5$ eV. On the right hand side, the configurations labeled Mo-T'(or T) for AB (or AA') stackings are energetically close, and the least stable.

We classify the relaxed configurations according to how the Mo$_\mathrm{imp}$ is related structurally to its neighboring sulfur atoms  \cite{kertesz1984octahedral,benavente2002intercalation}. The Mo-AB and Mo-H configurations form octahedral sites around the Mo$_\mathrm{imp}$. These configurations have a coordination number of six, corresponding to the six neighboring sulfur atoms. The Mo-T and Mo-T' configurations for the Mo$_\mathrm{imp}$ form a tetrahedral structure with a coordination number of four, therefore, the sulfur atoms in the top and bottom layers form a tetrahedral site for the Mo$_\mathrm{imp}$.
These octahedral and tetrahedral environments are shown schematically at the top of Fig. \ref{fig2}.
It is noteworthy that regardless of the final stacking, octahedral configurations are the most stable.

\subsection{Electronic Properties}

Results showing the band structures and the local density of states (LDOS) projected in space, for some of the considered configurations, are presented in Fig. \ref{fig3}.
We focus on the impurity in-gap states near the Fermi energy introduced by the Mo$_\mathrm{imp}$ atom, indicated by the areas enclosed by orange rectangles. The band structures of the most stable structures show three distinctive in-gap states, joined in two groups, labeled as regions \textbf{1} and \textbf{2} with degeneracies of one and two, respectively. Although the Mo$_\mathrm{imp}$ in these two configurations presents an octahedral sulfur environment, the in-gap bands present slightly different dispersive behavior.
In region \textbf{1}, the band for the Mo-AB case is more dispersive than the corresponding band for the Mo-H configuration, which is almost flat. The states in region \textbf{1} mainly have d$_{z^{2}}$ orbital character, as shown in the LDOS of panels (d) and (e).
Note that in region \textbf{1},  the surrounding region of the impurity for the Mo-AB case has some hybridization with bilayer orbitals,
which is not observed for the Mo-H case.
In region \textbf{2} of the Mo-AB case, there are two energy bands which are mainly non-bonding Mo$_\mathrm{imp}$ d orbitals with the neighboring sulfur atoms, as shown in the LDOS in panels (d) and (e).
By comparing the Mo-AB and Mo-H configurations, the stability order can be associated with the widening of the bands in regions \textbf{1} and \textbf{2} and to the displacement of the bands in region \textbf{2} to lower energies in the Mo-AB configuration.

The band structures in the Mo-AB and Mo-H configurations are spin-compensated and related to the impurity in an octahedral sulfur environment.
We now focus on the Mo-T and Mo-T' cases. These two configurations have similar energy band structures and are found to have close energies. Both systems exhibit spin polarized behavior with a total magnetic moment of 2 ${\mu_{B}}$, which is determined by a similar Mo$_\mathrm{imp}$ tetrahedral environment.
In particular for the Mo-T configuration, the spin up and spin down components of the spatial resolved LDOS are shown in Fig. \ref{fig3} panel (f).
The state in region \textbf{1} has a d$_{z^{2}}$ orbital character; however, it is above the Fermi level.
The LDOS of the lower impurity states, the up states in region \textbf{3} and the spin down
component in region \textbf{2} are nearly equal, depleting the LDOS in the
Mo$_\mathrm{imp}$-S bond direction as in the Mo-AB case.
%
%
However, the states responsible for the spin polarization in the T and T'
configurations are the spin up d-orbitals in region \textbf{2}, which are along the
Mo$_\mathrm{imp}$-S bonds.
We also find that the LDOS is localized not only on the Mo$_\mathrm{imp}$, but also on one of the MoS$_{2}$ layers.
This layer asymmetry indicates that doping by electrons or holes could spatially differentiate between the two
layers in the MoS$_{2}$ bilayer, a finding that could be of use in optoelectronic applications \cite{hong2014ultrafast,kosmider2013electronic}.

\begin{figure}[ht]
\centering
\includegraphics[width=\columnwidth]{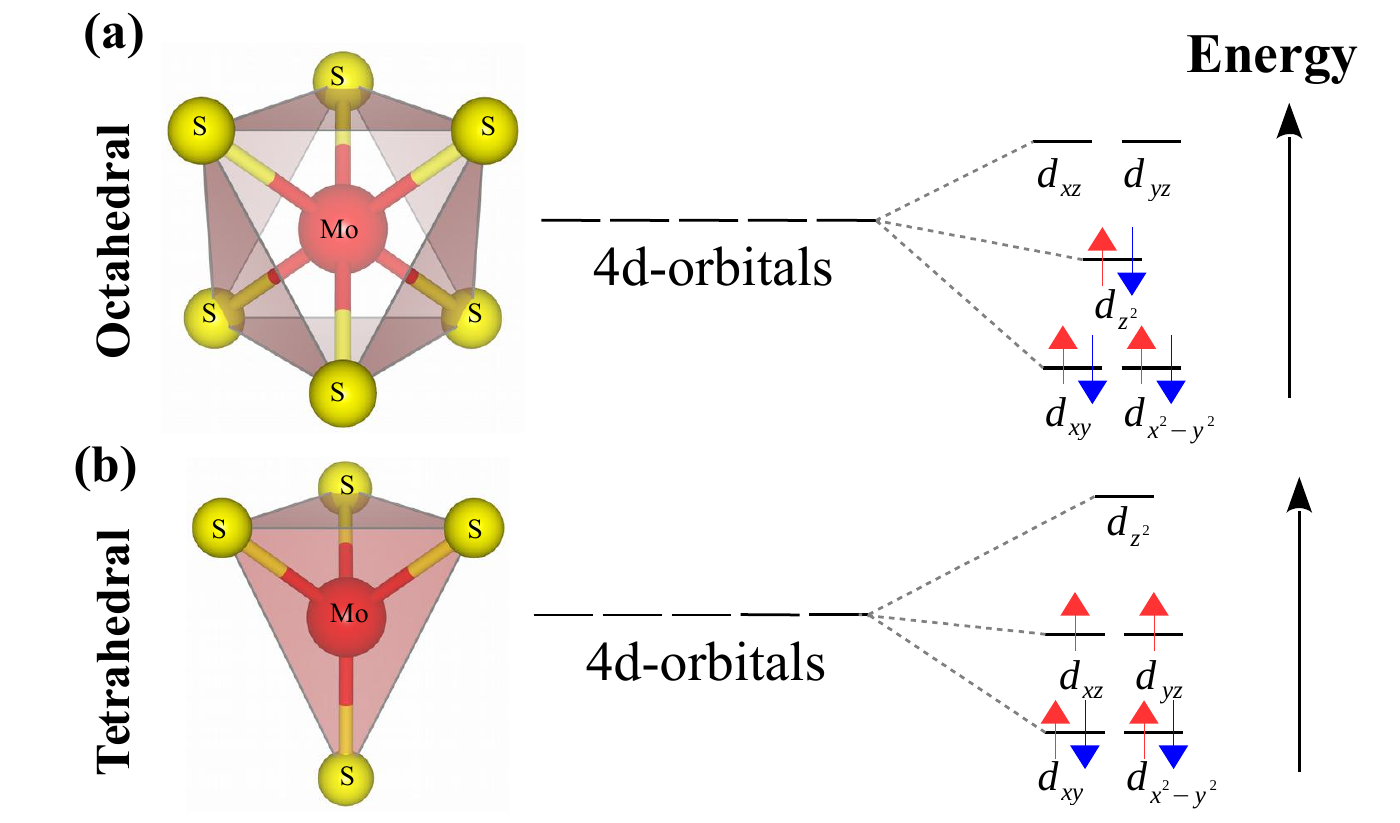}
\caption{Ligand field splitting for Mo$_\mathrm{imp}$ d-orbitals produced by the neighboring sulfur atoms in (a) octahedral and (b) tetrahedral environments. Figure prepared using VESTA \cite{vesta}.} \label{fig4}
\end{figure}

The electronic structure for the impurity level states is best understood  using crystal field theory. We analyze the ligand field splitting for the Mo$_\mathrm{imp}$ d-orbitals produced by the interactions with the sulfur ligands for the octahedral and tetrahedral sites. The bonding and non-bonding
interactions of d-orbitals for octahedral and tetrahedral sites are in agreement with the energy level scheme shown in Fig. \ref{fig4}.
We consider the \textit{z}-axis perpendicular to the layers and the \textit{x} and  \textit{y} axes in the in-plane layer.
In the octahedral environment for the Mo-AB and Mo-H configurations, the sulfur ligands overlap less with the in-plane d$_{xy}$ and
d$_{x^{2}-y^{2}}$ orbitals,  these orbitals are therefore non-bonding and have the lowest energy. The d$_{z^{2}}$ orbital
remains non-bonding at an intermediate energy, interacting less with the sulfur atoms. We next find that the d$_{xz}$ and d$_{yz}$ orbitals are more strongly directed and interact with the sulfur atoms along Mo$_\mathrm{imp}$-S bonds, lying at higher energies.
In the case of tetrahedral environment fo Mo-T' and Mo-T, shown in Fig. \ref{fig4}(b),
the d$_{xy}$ and d$_{x^{2}-y^{2}}$ orbitals behave similarly to the octahedral structure; however, the d$_{xz}$ and d$_{yz}$ orbitals exchange roles with the d$_{z^{2}}$ orbital. Thus, the d$_{z^{2}}$ orbital in the tetrahedral environment interacts more with the sulfur ligands increasing its energy, as shown for region \textbf{1} of Fig. \ref{fig3}(c) and (f) in the Mo-T case.
The level filling help to explain why the tetrahedral cases have spin polarization, and a total magnetization of 2 $\mu_B$, as shown by the arrow counting.

\begin{figure*}[ht]
\centering
\includegraphics[clip,width=\textwidth]{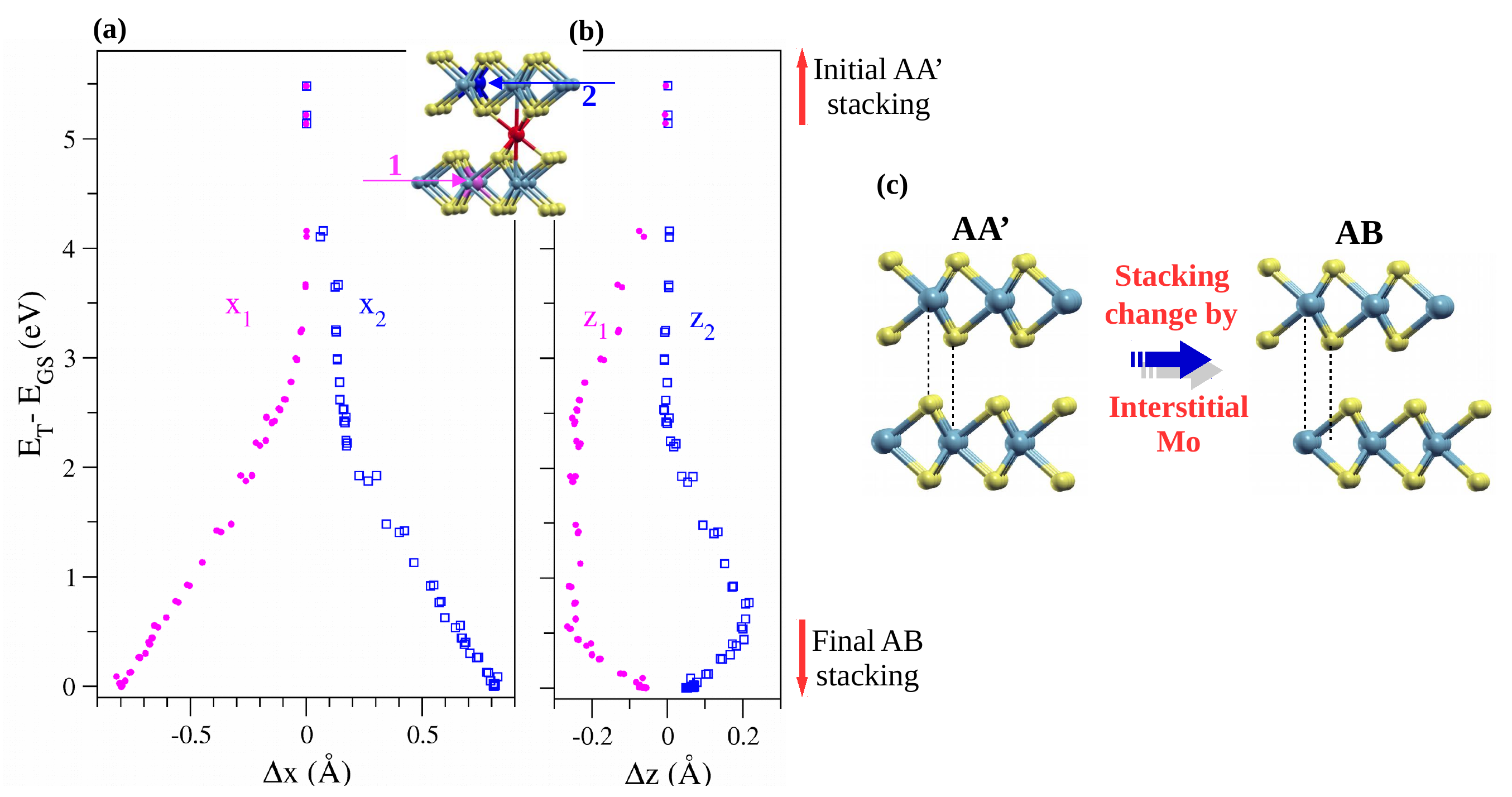}
\caption{(a-b) Energy change vs. relative displacement of two Mo atoms labeled as \textbf{1} in magenta and \textbf{2} in blue for Mo-AB configuration. E$_\mathrm{{T}}$ is the total energy in each relaxation step and E$_\mathrm{{GS}}$ is the ground state energy. The arrows indicate the final positions of the \textbf{1} and \textbf{2} atoms. x$_{1}$, x$_{2}$ and z$_{1}$, z$_{2}$ are the displacements in the \textit{x}-direction and the \textit{z}-direction, for atoms \textbf{1}, \textbf{2}, respectively. The zero point displacements in the \textit{x}-direction and \textit{z}-directions are set for AA' stacking. (c) Scheme showing the change from the initial AA' to the relaxed AB stacking induced by the Mo impurity.} \label{fig5}
\end{figure*}

We now consider the gap changes in the bulk bands, shown in gray in Fig. \ref{fig3}, and induced by the Mo$_\mathrm{imp}$.
It is well known that band gaps calculated using GGA
infra-estimate the values produced in experiments, so we discuss differences in magnitude gaps. The layer-gap is indicated by the energy difference between the HOMO-LUMO bulk bands at the $\Gamma$-point
\footnote{Note that the two-gap scenario of the original $1 \times 1$ unit cell is reduced to a direct gap at the $\Gamma$-point due to k-space folding.}.
The energy gaps of the pristine MoS$_2$ layers are correlated with the interlayers distances. We check that the gap and distances for the AB pristine
stacking are $0.09$ eV smaller and $0.03$ \AA{} shorter than the values
for the AA' pristine case.
The layer-gaps and the interlayer separation including Mo$_\mathrm{imp}$ show larger values in comparison with the pristine cases.
Among the Mo doped systems, the most stable Mo-AB case has the smaller layer-gap and the shortest interlayer separation. The layer-gap is $0.2$ eV above the AB-pristine, and the interlayer separation is $0.03$ \AA{} larger than that of AB-pristine.
The layer-gap values for the Mo-H and the T and T’ increase from the Mo-AB case by $0.2 $ eV and by $0.5$ eV respectively.
These gap differences are somewhat correlated with the difference between
layer-layer distances, $\sim$ 0.5 \AA, between the Mo-AB and Mo-H cases,
a value that increases up to $0.7$ \AA{} for the T(T') configurations.
The increase in the layer-band-gap with interlayer distance is explained
by a weaker interlayer coupling.

The interlayer distances in the proximity of the impurity are between
$0.1$ and $0.16$ \AA{} larger than those far from it, which indicates
the role of local strain.
Furthermore, experiments prove than the band gap of bilayer TMDCs can
be controlled by strain 
\cite{castellanos2013local,san2016inverse,feng2012strain}.
We propose that the electronic and structural modifications around the
impurity could be used in a similar way to electronic confinement for embedded
quantum dots. Current experimental techniques employing cross-sectional
scanning transmission electron microscope analysis in
encapsulated TMDC materials can provide evidence of impurity species being
trapped in the interstitial region \cite{rooney2017observing}.  This effect
thus has potential applications for optoelectronic devices as exciton traps
around Mo-doped bilayers.
A number of different experimental techniques can be used to corroborate our theoretical predictions, for instance angle-resolved photoemission spectroscopy and cross-sectional scanning transmission electron microscope analysis \cite{brauer1999modifying,rooney2017observing}.

\subsection{Stacking change}
Another possible implication of our results is that transition metal ions could be used to engineer the stacking between TMDC bilayers and to tune their electronic properties.
In the Mo-AB and Mo-H configurations, the Mo$_\mathrm{imp}$ is located within sulfur ligands forming octahedral sites. In these two configurations, the Mo$_\mathrm{imp}$ presents structural differences in the relative position respect to the nearest Mo atoms, belonging to the top and bottom MoS$_{2}$ layers.
%
The Mo$_\mathrm{imp}$ bonding produces a stacking change in Mo-AB related to the total energy decrease.
The interlayer Mo-Mo distance around the impurity is smaller around 0.1 \AA{} than the interlayer distance away from the impurity.
The shorter distance promotes the hybridization of the impurity states with the layer states increasing the dispersion of the in-gap impurity states.

Figure \ref{fig5} (a-b) shows the total energy change versus the relative displacement in the \textit{x} and \textit{z} directions, for two distant Mo atoms during the relaxation steps, starting with the AA' stacking. We find that the total energy drops until it reaches the reference ground state energy.
The displacements in the \textit{z}-direction show that the layers are breathing towards the ground-state AB stacking.
The energy at the maximum turning point is $\sim 0.5$ eV, a value that reflects the barrier to be overcome by shifting the layers.
In Fig. \ref{fig5}(c), we schematize the stacking change from AA' to AB mediated by the presence of a Mo$_\mathrm{imp}$ atom, the stability gain produced by the Mo$_\mathrm{imp}$ favors AB stacking.
DFT calculations controlling the sliding of pristine bilayers yield a similar
energetic profile, but in the reverse direction to the Mo-doped bilayers \cite{peng2014stacking}.


\section{Final Remarks}
We studied the structural and electronic properties of a MoS$_{2}$ bilayer with intrinsic Mo impurities within the interlayer region.
We find that the most stable con\-fi\-gu\-ra\-tion is Mo-AB, with an energy gain above the van der Waals interaction because the Mo impurity levels strongly hybridize with the nearest atoms.
Note that pristine bilayers have AA' stacking as the ground state.
A change in the stacking stability order from AA' to AB is observed to be induced by impurities, with the corresponding variation in energy gap. Thus, it is possible to engineer the stacking between TMDC bilayers during the growth process, enabling their electronic properties to be fine-tuned.
The states and deformations induced by impurities could also be used for electronic confinement applications in optoelectronic devices, based on exciton/electron trapping.
%


\section*{Acknowledgments}

This work was part financed by a Fondecyt grant 1140388 and Anillo Bicentenario de Ciencia y Tecnolog\'ia, Conicyt grant Act-1204. J.W. Gonz\'alez and A. Ayuela acknowledge the financial support of the  Spanish  Ministry  of  Economy  and  Competitiveness  MINECO projects  FIS2013-48286-C02-01-P  and  FIS2016-76617-P,  the  Basque  Government  under the ELKARTEK project(SUPER), and the University of the Basque Country grant No.  IT-756-13.
N. Cort\'es acknowledge support from the Conicyt grant, No 21160844 and the hospitality of CFM-MPC and DIPC.
The authors are indebted to Prof. B. Harmon, L. Chico and T. Alonso-Lanza for their helpful discussions, we also acknowledge the technical support of the DIPC computer center.

\bibliography{bib}
\email{natalia.cortesm@usm.cl}

\end{document}